\documentclass[preprint2]{aastex}

\def\J{{\rm Je}}
\def\K{{\rm Kn}}
\def\M{{\rm Ma}}
\def\Me{{\rm M_{\rm escape}}}
\def\Mp{{\rm M_{\rm Parker}}}
\def\Mj{{\rm M_{\rm Jeans}}}
\def\Mf{{\rm M_{\rm Fourier}}}
\def\Khs{{\rm Kn_{\rm hs}}}

\def\be{\begin{equation}}
\def\ee{\end{equation}}

\begin{document}

\title{The rate of thermal atmospheric escape.} 

\author{Andrei Gruzinov}

\affil{CCPP, Physics Department, New York University, 4 Washington Place, New York, NY 10003}

\begin{abstract}

A formula is derived for the rate of thermal atmospheric escape, valid, and asymptotically exact, at low Knudsen number. 

~~

\end{abstract}

\section{The thermal escape problem}

\begin{figure}
\plotone{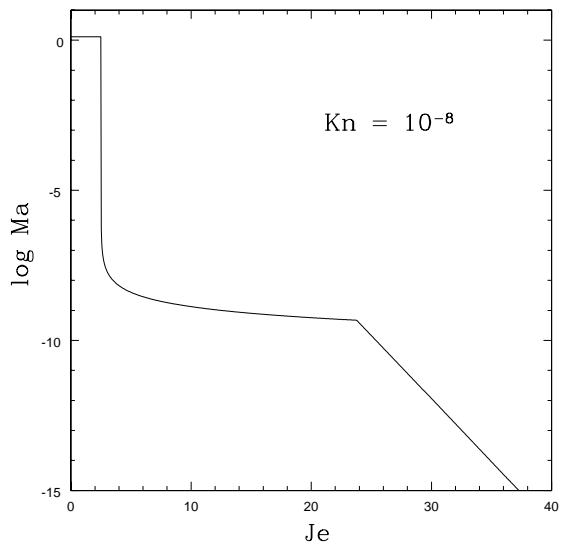}
\caption{Escape rate for $\K =10^{-8}$, $c_p=5/2$, $\alpha =1$. $\J <2.5$ is the Parker regime, $2.5<\J <24$ is the Fourier regime, $\J >24$ is the Jeans regime.}
\end{figure}

Consider a planetary atmosphere without stirring and radiative heating or cooling above the nominal surface. The atmosphere cannot be static since a static atmosphere reaches thermodynamic equilibrium. The Boltzmann distribution would then give a finite density at infinite distances, which is unphysical.

The atmosphere must be in a state of permanent escape. We want to calculate the rate of escape and the resulting temperature and density profiles.

The thermal escape problem has a long history (Jeans 1904, Parker 1958, Hunten 1982). But in recent applications to planetary bodies, there has been disagreement on which model -- hydrodynamic or free molecular flow -- should be applied (Johnson 2010). We believe that all issues have been finally settled by a direct molecular dynamics simulations of Volkov et. al. (2010). Here we just show how the Volkov et. al. (2010) results can be obtained analytically.

It would seem that the escape rate calculation is only possible via molecular dynamics or Boltzmann numerical simulation, because the exobase region, where the transition from hydrodynamic to free molecular flow occurs, cannot be treated analytically. But we show, for low Knudsen number at the surface, that the exobase boundary conditions are either irrelevant or can be deduced due to large overlap between hydrodynamic and free molecular flow regions. This allows one to calculate the atmospheric escape using hydrodynamics and to derive a formula for the escape rate.

For clarity, we approximate the molecular interaction potential by a power law (this includes hard-sphere gas as a limiting case). Then, by scaling arguments, the escape rate must be given by
\be
\M =\Me (\J ,\K ).
\ee
Here we measure the escape rate by the radial velocity at the surface $v$, and the velocity is represented by the isothermal Mach number
\be
\M \equiv {v \over \sqrt{ T/m} }.
\ee
The Jeans number is 
\be
\J \equiv {GMm\over TR}, 
\ee
where $G$ is the gravitational constant, $M$ is the planet mass, $m$ is the molecule mass, $T$ is the surface temperature, $R$ is the surface radius. The Knudsen number is $\K \sim {\lambda \over R}$, where $\lambda$ is the mean free path. We define
\be
\K \equiv {m\kappa \over R\rho \sqrt{ T/m}},
\ee
where $\kappa$ is the thermal conductivity, $\rho$ is the density.

The escape rate formula is given in \S2. The formula includes three branches. In \S3-5, we give the recipes for calculating the branches. In \S6, we explain the origin of the recipes, describe numerical hydrodynamics which confirms the escape rate formula, and compare our results to the molecular dynamics simulations of Volkov et. al. (2010).

\section{The escape rate formula} 

In the low-Knudsen limit, 

\be \label{form}
 \Me = \left\{ \begin{array}{rl}
\Mp (\J ), & \J < c_p; \\
\Mf (\J, \K ), & c_p < \J \lesssim \ln (\K ^{-1}); \\
\Mj (\J ), & \J \gtrsim \ln (\K ^{-1}).
\end{array}\right.
\ee
Here $c_p$ is the heat capacity. The formula is asymptotically exact (Appendix). 

$\Mp$ is just the Parker wind (Parker 1958) escape rate. It depends on $\J $ and $c_p$. For example, for $c_p=5/2$, 
\be
\Mp =\sqrt{5/3}.
\ee
Formula (\ref{form}) limits the $\J$ range over which the Parker rate applies. The Parker regime is described in \S3.

$\Mj$ is the Jeans (1904) escape rate. It depends on $\J $ only:  
\be
\Mj={(\J +1)\exp( -\J )\over \sqrt{2\pi } }
\ee
Formula (\ref{form}) extends the $\K$ range over which the Jeans rate applies. The Jeans regime is described in \S4.

$\Mf$ is proportional to $\K$, with the $\J$-dependent coefficient of proportionality: 
\be
\Mf (\J ,\K )=F(\J )\K.
\ee
Here $F(\J )$ depends on $\J $, $c_p$, and 
\be
\alpha \equiv {d\ln \kappa \over d \ln T}.
\ee
For example, for $\alpha =1$, 
\be
F(\J )={1\over \J -c_p}
\ee
For uniformity of notation, we call it the Fourier regime. The Fourier regime is described in \S5.

To illustrate the formula, Fig. 1 shows the escape rate as a function of $\J$, for $\K =10^{-8}$, $\alpha =1$, and $c_p=5/2$. The different regimes were extended to the intersection points. Fig.1 is in qualitative agreement with molecular dynamics results of Volkov et. al. (2010). Quantitative comparison is done in \S6, where we also show that a simplified dissipative hydrodynamics model reproduces the molecular dynamics results to a good accuracy.

In the next three sections we calculate the escape rate postulating the properties of the flow. The assumptions made can be shown to be self-consistent. But, in fact, what makes the results trustworthy is that they do match the numerical hydrodynamics results described in \S6, and they also match the molecular dynamics simulations of Volkov et al (2010). 

The Parker regime is a high-velocity ideal flow with negligible thermal conductivity. The Fourier regime is a flow with algebraically small velocity and with thermal conduction balancing the ideal-hydrodynamic energy flux. The Jeans regime is a flow with exponentially small velocity, such that thermal conduction dominates.

\section{Parker regime} 

For $\J < c_p$, $\K \ll 1$ one can use ideal hydrodynamics \footnote{ More precisely, dissipative effects must become negligible after a possible boundary layer is passed. Whether the boundary layer is present, depends on how the atmosphere is fed. If the boundary layer is present, we simply move the nominal surface up, beyond the boundary layer.}. To write equations in the simplest form, we use scalings to set $M=m=R=1$, and also $\rho =1$, $T=1$, at the surface $R=1$. Now we need to calculate the outflow velocity at the surface $v_0=\Mp$, for any $\J =G<c_p$. 

We write stationary ideal hydrodynamics as flux conservation, energy flux conservation, and adiabaticity
\be \label{p1}
\begin{array}{l}
\rho v r^2=v_0,\\
{v^2\over 2}+c_pT-{G\over r} = {v_0^2\over 2}+c_p-G,\\
\rho =T^{c_p-1}.
\end{array}
\ee

From eqs.(\ref{p1})
\be \label{p2}
{v^2\over 2}+c_p\left( {v_0\over r^2v} \right) ^{{1\over c_p-1}}={v_0^2\over 2}+c_p-G+{G\over r}.
\ee
This ``algebraic'' equation for $v$ must have solutions for all $r>1$, such that the corresponding density $\rho =v_0/(r^2v)$ goes to zero at large $r$. This requirement limits the range of possible values of $v_0$.

For $c_p<3$, and also for $c_p>3$, $G<2c_p/(c_p-1)$, the smallest possible value is $v_0=\sqrt{c_p/(c_p-1)}$, in which case the sonic point is at the surface. For $c_p>3$ and larger $G$, $G>2c_p/(c_p-1)$, the sonic point moves out to $r>1$, and $v_0$ decreases. 

It is the smallest possible $v_0$ that we call $\Mp$. Time-dependent ideal hydrodynamics simulations and simulations of Volkov et al (2010) confirm that the flow does approach the smallest $v_0$ allowed by stationary ideal hydrodynamics. 

The above results can be written as
\be
\Mp=\sqrt{{c_p\over c_p-1}},
\ee
for $c_p<3$, and also for $c_p>3$, $\J <2c_p/(c_p-1)$. For $c_p>3$, $2c_p/(c_p-1)<\J <c_p$, $\Mp$ decreases, reaching zero at $\J =c_p$. The precise value of $\Mp$ is then given implicitly by an ``algebraic'' equation which can be derived from eq.(\ref{p2})\footnote{ $\left( {\M ^2\over 2}+c_p-\J \right) ^{c_p-{5\over 2}}=\left( {c_p\over c_p-1}\right) ^{c_p-1}(c_p-{5\over 2})^{c_p-{5\over 2}}{4\M \over \J ^2}$.}.

\section{Jeans regime} 

Fix the Knudsen number and let the Jeans number increase. Then the  escape rate approaches 
\be \label{j1}
\Mj={(\J +1)\exp( -\J )\over \sqrt{2\pi } }
\ee
-- the standard Jeans rate -- density normalized flux of positive-energy positive-radial-velocity molecules, calculated for Maxwellian distribution.

Eq.(\ref{j1}) is an exponentially accurate approximation (9\% error at $\J=2.5$, 0.008\% error at $\J =10$) of the true escape rate for collisionless molecular outflow with Maxwellian injection at the surface, which in our notations is 
\be
\Mj = {\int _0^\infty udu \int _{-\infty }^\infty dv fv \over \int _0^\infty udu \int _{-\infty}^\infty dv f}.
\ee
Here $v$ is the radial velocity, $u$ is the absolute value of the tangential velocity, $f\propto \theta \exp (-(u^2+v^2)/2)$ is the truncated Maxwellian. $\theta =0$ if simultaneously the energy is positive, $(u^2+v^2)/2-\J >0$, and the radial velocity $v$ is negative; otherwise $\theta =1$. 

\section{Fourier regime} 

Fix the Jeans number $\J >c_p$ and let the Knudsen number decrease. The escape rate then approaches 
\be \label{f1}
\Mf =F(\J )\K .
\ee
The coefficient $F(\J )$ is calculated below.

In the Fourier regime, the negative hydrodynamic energy flux carried by the flow is almost exactly compensated by the thermal conductivity: 
\be \label{q1}
r^2\kappa T'=\Phi (c_pT-{G\over r}).
\ee
Here $\Phi=\rho v r^2=$const is the flux per steradian, $M=m=1$ was set by scaling, prime denotes the $r$-derivative, kinetic energy and viscosity are negligibly small.

Using scalings, we put $T=1$ at $r=1$, and then solve eq.(\ref{q1}) for $r>1$. We adjust $\Phi$ in such a way as to get the temperature going to zero at large radii (otherwise the flow either stalls or becomes singular at finite radii, never reaching an exobase, \S6). We want to solve the problem for arbitrary $\alpha \equiv d\ln \kappa /d\ln T$ \footnote{$\alpha =1$ is a good approximation for real gases at relevant temperatures; $\alpha =0.5$ for the hard-sphere gas, and we want to reproduce the results of Volkov et. al. (2010)}.

First consider just the case $\alpha =1$. One can check numerically or analytically, that there is only one solution of eq.(\ref{q1}) with correct asymptotic behavior at large $r$:
\be 
T={1\over r}
\ee
Substitution into eq.(\ref{q1}) then gives
\be 
{\kappa \over T}=(G-c_p)\Phi.
\ee
Or, in different notations,
\be 
F(\J ) = {1 \over \J -c_p }.
\ee

For arbitrary $\alpha <1$, numerical integration of eq.(\ref{q1}) shows that there is still a unique choice of $\Phi$ which gives a non-singular zero-at-infinity temperature. One can show that for $0<\J-c_p\ll 1$,
\be 
F={1\over \J -c_p}.
\ee
while for $\J \gg 1$, 
\be 
F={2\over 1+\alpha} {1\over \J }.
\ee
Joining the asymptotics as follows
\be \label{f2}
F(\J )={1\over (\J -c_p) f(\J )}
\ee
\be \label{f3}
f(\J ) \approx 1-{1-\alpha \over 2}{\J -c_p\over \J -0.55c_p}.
\ee
gives an error less than $0.3\%$ for all relevant $\alpha$, $c_p$ and $\J$.

\section{Hydrodynamics} 

To show that the escape rate formula (\ref{form}) is asymptotically exact at $\K \rightarrow 0$, and to improve the accuracy in the non-asymptotic regime, we use stationary non-ideal hydrodynamics. The flux conservation, the energy flux conservation, and the momentum equations are
\be \label{h}
\begin{array}{l}
\Phi =\rho vr^2, \\
Q=\Phi ({v^2\over 2} +c_pT-{G\over r}) -{4\over 3}\eta r^2v(v'-{v\over r})-r^2\kappa T', \\
\rho vv'=-(\rho T)'-{G\rho \over r^2}+{4\over 3}{1\over r^3}(r^3\eta (v'-{v\over r}))'.
\end{array}
\ee
Here $\Phi$, $Q$ are constant fluxes, $\eta$ is the viscosity\footnote{ We have assumed that the second viscosity vanishes. This is true for the hard-sphere gas and allows precise comparison to the results of Volkov et. al. For polyatomic gases, the second viscosity must be included, but it turns out that the viscosity effects are negligible at low Knudsen, unless $\J$ is very close to $c_p$.}. The boundary conditions at $r=1$ are
\be
\rho =1, T=1, v=\Phi
\ee

Given $G =\J$ and $\kappa (T=1) = \K$, we must find a pair $(\Phi ,Q)$ that gives the ``true'' flow \footnote{ To solve (\ref{h}) we also need $v'(1)$. But this boundary condition is irrelevant. Wrong $v'(1)$ would result in a boundary layer, and we have agreed to place the nominal surface above the boundary layer.}.  Which flows are ``true'' is decided by the boundary conditions at the exobase, and it would appear that one cannot proceed without a molecular dynamics simulation of the exobase region. 

But it turns out that for $\K \ll 1$ the mere requirement that the exobase exists allows to calculate the escape flow in the Parker and Fourier regimes. In the Jeans regime, this trick no longer works -- there are many outflows with an exobase. What allows to solve the problem in the Jeans regime is a large overlap between hydrodynamic and free molecular flow regions.

{\it Parker regime --} Take $G<c_p$, and, as always, $\kappa =\K \ll 1$. If the initial velocity is large, $v_0\equiv \Phi \sim 1$, the non-ideal hydrodynamics (\ref{h}) degenerates into the ideal hydrodynamics (\ref{p1}), and we get the Parker escape.

Only for small velocities, $v_0\lesssim \kappa$, dissipative effects can be significant. But numerical integration of (\ref{h}) shows that for small $v_0$ the exobase is never achieved. $\K$ stays small at all radii, no matter what energy flux $Q$ one chooses.

What sets the boundary of the Parker regime to $\J =c_p$ is the sign of the ideal hydrodynamic energy flux in the small-velocity limit
\be
Q_{\rm h}=\Phi (c_pT-{G\over r}).
\ee
For $\J <c_p$ this is positive. Now we show that this means no exobase for small $v_0$.

For small $v_0$, the density is approximately given by the equilibrium condition
\be \label{eq}
-(\rho T)'-{G\rho \over r^2}=0.
\ee
At moderate $G$, the only way to reduce the density enough to get an exobase,  is by decreasing the temperature at large distances. If the temperature remains finite at infinity, the density saturates as 
\be
\rho \propto \exp({G\over Tr}). 
\ee

At small $v_0$, we can drop the $v^2$ terms from the energy flux expression in (\ref{h}), and get 
\be \label{hc}
r^2\kappa T'=\Phi (c_pT-{G\over r})-Q.
\ee
For the temperature to decrease, we must take a positive flux $Q$. But finite $Q$ then dominates the r.h.s. of eq.(\ref{hc}) at large radii, and we get
\be 
r^2\kappa T'=-Q.
\ee
This gives
\be \label{tem}
T\propto r^{-{1\over 1+\alpha }}.
\ee
This temperature decreases -- but too slowly. For $\alpha >0$, eqs.(\ref{eq}, \ref{tem}) give a density growing at infinity. 

In summary, small-velocity flows have no exobase, and large-velocity flows are ideal, giving the Parker flow. The lack of collisions beyond the exobase cannot change the escape rate of the highly supersonic outflow.

{\it Fourier regime --} At fixed $\J >c_p$, as $\K$ decreases, the set of pairs $(\Phi ,Q)$ which give solutions with an exobase shrinks to a single point $(\Mf ,0)$. If $(\Phi ,Q)\neq (\Mf ,0)$, the flow either becomes singular at a finite radius or goes to infinite radii, without ever reaching local $\K \sim 1$. So in this case too, we can claim that $\Mf$ is an exact asymptotic even without knowing what happens at the exobase. 

The only way to reach an exobase is to set $Q=0$ in eq.(\ref{hc}), and then tune $\Phi$, as described in \S5. 

{\it Jeans regime --} Fix $\K$ and let $\J$ increase. The outflow velocity becomes so small that thermal conduction in eq.(\ref{hc}) dominates and we get a nearly constant temperature profile. Eq.(\ref{eq}) then gives an exponential density. These temperature and density are also predicted by the free molecular outflow at high $\J$. We see that the two descriptions agree.

We can therefore use hydrodynamic description well above the point $r=R_e$ where the density scale height becomes comparable to the mean free path, $\K (R_e) = \J ^{-1}(R_e)$. But for $\K (r) \gg \J ^{-1}(r)$, we can also use the free molecular outflow results to calculate the fluxes $(\Phi ,Q)$. Now, we use hydrodynamics to integrate back to $r=1$. One can show, or just check numerically, that this procedure, for $\J \gg \ln \K ^{-1}$, gives the Jeans rate (to leading order in $\J$ in the prefactor).

{\it Simplified hydrodynamics --} Both the Fourier and the Jeans regimes can be described simultaneously, in a much simplified version of hydrodynamics which we give below. One can then hope that even between the asymptotics this hydrodynamic description will work (it does work for hard spheres, see below).

The model has one adjustable parameter $\K _0\sim 1$. We set $T=1$, $\rho=1$, $\K =\K _0$ at $r=1$, and use the free molecular flow to calculate $\Phi$ and $Q$ at $r=1$:
\be \label{sh}
\Phi={e^{-G}(G+1)\over \sqrt{2\pi }},~Q={e^{-G}((c_p-{3\over 2})G+c_p-{1\over 2})\over \sqrt{2\pi }}.
\ee
We then integrate eq.(\ref{hc}) backwards in radius, to $r<1$. Simultaneously we integrate eq.(\ref{eq}) and calculate $\K (r)$. The integration is terminated when the desired value of $\K$ is reached. We adjust $G$ so that the desired value of $\J$ is reached simultaneously with the desired value of $\K$. This gives the escape rate for these $\K$ and $\J$. One also gets correct temperature and density profiles for $G\gg 1$ and also for $r\ll 1$ at any $G$. We have checked, by numerical integration of the full system (\ref{h}), that neglecting viscosity and inertia is justified, again for $G\gg 1$ and also for $r\ll 1$ at any $G$.

We have used the simplified hydrodynamics (\ref{hc},\ref{sh}) to establish the logarithmic accuracy of the escape rate formula (\ref{form}) in the Fourier-Jeans transition region (Appendix).

{\it Comparison with Volkov et. al. (2010) --} Now compare the predictions of our escape rate formula eq.(\ref{form}) to the molecular dynamics simulations of Volkov et. al. (2010). We put $c_p=5/2$, $\alpha =1/2$. We also remember that our definition of $\K$ is different by a constant factor from the conventional hard-sphere Knudsen $\Khs=0.34\K$. 

We find that for $\J=3, 10, 15$, $\Khs =10^{-2}, 10^{-3}, 10^{-4}$ the simplified hydrodynamics (\ref{hc},\ref{sh}) (with $\K _0=3$) predicts the escape rates which agree with the results of Volkov et. al. (2010) to about 10\% for all points but one. 

Let us discuss this discrepancy. For $\J=3$, the escape rate formula (\ref{form}) gives for $\Khs =10^{-2}, 10^{-3}, 10^{-4}$ (using eqs.(\ref{f1}, \ref{f2}, \ref{f3})) 
\be 
{\M \over \Khs } =6.4,~6.4,~6.4
\ee
The simplified hydrodynamics (\ref{sh}) gives 
\be 
{\M \over \Khs } =5.6,~6.0,~6.2
\ee
Fig.4 of Volkov et al (2010) gives 
\be 
{\M \over \Khs } =5.0,~6.8,~4.1
\ee
The simplified hydrodynamics works as expected -- as we go deeper into the Fourier regime, predicted $\M$ approaches $\Mf$. The first two values of Volkov et. al. are also reasonable -- the error of $\Mf$ decreases as we get deeper into the Fourier regime. But the last point should have been even closer to the Fourier value of 6.4. Some 50\% error is unexpected. 

We believe that it is the molecular dynamics results at $\J =3$, $\Khs =10^{-4}$ which have a large error. We have checked, by integrating the full system (\ref{h}) numerically, that a hydrodynamic flow, starting at $\Khs =10^{-4}$, $\M  =4.1\times 10^{-4}$, $\J =3$, never goes above $\Khs =1.5\times 10^{-3}$ for any $Q$. It seems unlikely that molecular dynamics can noticeably deviate from hydrodynamics at such low Knudsen numbers.

\section{Summary} 

Formula (\ref{form}) gives the thermal escape rate at low Knudsen.

The escape occurs either in Parker, in Fourier, or in the Jeans regime. In the Parker regime, we get a supersonic ideal hydrodynamical wind. The temperature and density profiles are given by the Parker wind model (\S3). Both temperature and density decrease algebraically. In the Jeans regime, we have a nearly isothermal slowly escaping atmosphere, with exponential density. In the Fourier regime, thermal conductivity balances the ideal energy flow (\S5). The temperature and the density decrease algebraically.

\acknowledgements

I thank Robert Johnson for useful discussions.  

\vskip 0.3cm

{\it Appendix} 

The escape rate formula (5) is asymptotically exact in the following sense:
\be
\J ={\rm const}<c_p, \K \rightarrow 0: {\Me \over \Mp } \rightarrow 1,
\ee
\be
\J ={\rm const}>c_p, \K \rightarrow 0: {\Me \over \Mf } \rightarrow 1,
\ee
\be
\K ={\rm const}, \J \rightarrow \infty : {\Me \over \Mj } \rightarrow 1,
\ee
\be
\J >c_p, \K \rightarrow 0: {\ln (\Me ) \over \ln ({\rm min}(\Mf ,\Mj ))} \rightarrow 1.
\ee

In words: (i) As Knudsen decreases at fixed Jeans, the Parker and Fourier formulas approach the true escape rate everywhere but near the Parker-Fourier transition.  As Knudsen decreases, the Jeans number width of the Parker-Fourier transition region shrinks to zero. (ii) As Jeans increases at fixed Knudsen, the Jeans formula approaches the true escape rate. (iii) In the Fourier-Jeans transition interval, both formulas are valid, but only with logarithmic accuracy.


\begin{references}



\reference{} Jeans, J. H., 1904, ``The dynamical theory of gases'', Chapter XVII, University Press, Cambridge.

\reference{} Parker, E. N., 1958, ApJ, 128, 664

\reference{} Hunten, D. M., 1982,  Planet. Space Sci., 30, 773

\reference{} Johnson, R. E., 2010, ApJ, 716, 1573

\reference{} Volkov, A. N., et al,  2010,  arXiv:1009.5110

\end{references}
\end{document}